\documentclass{iau} 
\usepackage{graphicx}
\usepackage{amsmath}
\usepackage{graphicx,caption,subcaption}
\usepackage{aas_macros}

\title[] %% give here short title %%
{Magnetic field topology from non-force free extrapolation and magnetohydrodynamic simulation of its eventual dynamics}

\author[ Sushree S. Nayak, R. Bhattacharyya, A. Prasad \& Qiang Hu]   %% give here short author list %%
{Sushree S. Nayak$^1$, R. Bhattacharyya$^1$, A. Prasad$^1$
 \& Q. Hu$^2$}

\affiliation{$^1$Udaipur Solar Observatory, Physical Research Laboratory, Dewali, \\ Bari Road, Udaipur-313001, India \\ email: {\tt sangeetan@prl.res.in}, email: {\tt ramit@prl.res.in}, email: {\tt avijeet@prl.res.in}\\
$^2$Center for Space Plasma and Aeronomic Research, The University of Alabama in Huntsville, Huntsville, AL 35899, USA  \\ email: {\tt qh0001@uah.edu}}

\pubyear{2018}
\volume{340}  %% insert here IAU Symposium No.
\jname{Long-Term Datasets for the Understanding of Solar and Stellar Magnetic Cycles}
\editors{D. Banerjee, J. Jiang, K. Kusano  \& S. Solanki}
\begin{document}

\maketitle

\begin{abstract}
Magnetic reconnections (MRs) for various magnetic field line (MFL) topologies
are believed to be the initiators of various solar eruptive events like
flares and coronal mass ejections (CMEs). Consequently, important is a thorough
understanding and quantification of the MFL topology and their evolution which leads to
MRs. Contemporary standard is to extrapolate the coronal MFLs using equilibrium models
where the Lorentz force on the coronal plasma is zero everywhere, because either there is
no current or the current is parallel to the magnetic field. In tandem, a non-force-free-field
(NFFF) extrapolation scheme has evolved and allows for a Lorentz force which is non-zero only at the photosphere but asymptotically vanishes with height. The paper reports magnetohydrodynamic (MHD)-
simulations initiated by NFFF extrapolation of the coronal MFLs for the active region
NOAA 11158. Interestingly, quasi-separatrix layers (QSLs) which facilitate MRs are
detected in the extrapolated MFLs. The AR 11158 is flare producing and, the paper makes
an attempt to asses the role of QSLs in the flare onsets.
\end{abstract}

\firstsection % if your document starts with a section,
\section{Introduction}
The phenomenon of MR: a process where MFLs rearrange while magnetic energy gets converted into heat and kinetic energy of the plasma, is believed to initiate flares and CMEs. Occurrence of MRs strongly depends on the MFL topology and hence it is imperative to explore their interdependence, the focus of the paper.
For the purpose, we select the well documented active region NOAA 11158 and extrapolate MFLs by using magnetograms from HMI/SDO at 01:12 UT on 15th February, 2011, which is roughly half an hour before the peak of an X2.2 class flare (1:45 UT) followed by an earth-directed CME. The coronal MFLs are constructed by NFFF extrapolation technique (\cite[Hu et al. 2008]{2008ApJ...679..848H}), resolving a physical domain of extents $\approx$ $268\times134\times134$ (in Mm) 
by a computational domain having 
$128\times64\times64$ grids in the $x$, $y$ and $z$ respectively. The normalized deviation of the extrapolated field at the photosphere from its magnetogram value: En=0.227; which renders the extrapolation reasonably accurate. Figure (1) overlay extrapolated MFLs on the AR image in 94 {\AA} channel. Evident is the overall good match between the MFLs and the coronal loops. Importantly, Quasi Separatrix Layers (QSLs) regions having large gradient of MFL connectivity are potential sites for MRs 
(\cite[D{\'e}moulin 2006]{2006AdSpR..37.1269D}). QSLs are also found to be present whereas magnetic flux ropes are completely absent.  The presence of QSLs are further confirmed by large values of the squashing factor (not shown). To explore such MRs, MHD simulations are performed using the parallelized three dimensional numerical model EULAG-MHD. The computations are carried out on the Vikram-100, the 100TF computational facility at the Physical Research Laboratory.

\begin{figure}[ht]
\centering
\includegraphics[width=.36\linewidth]{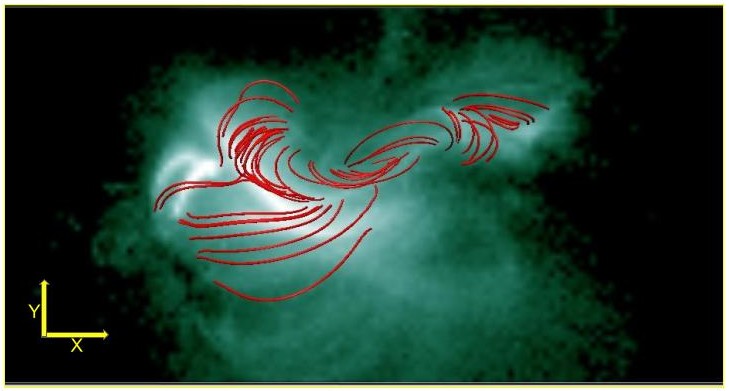}
\caption{ Extrapolated MFLs plotted over 94 {\AA}  image from AIA/SDO for AR
11158 at 01:12 UT. The dimension of the image is ($128\times64$) respectively.}
\end{figure}

\begin{figure}[ht]
  \centering
  \begin{subfigure}[]{0.45\textwidth}
    \centering
    \includegraphics[width=0.95\linewidth]{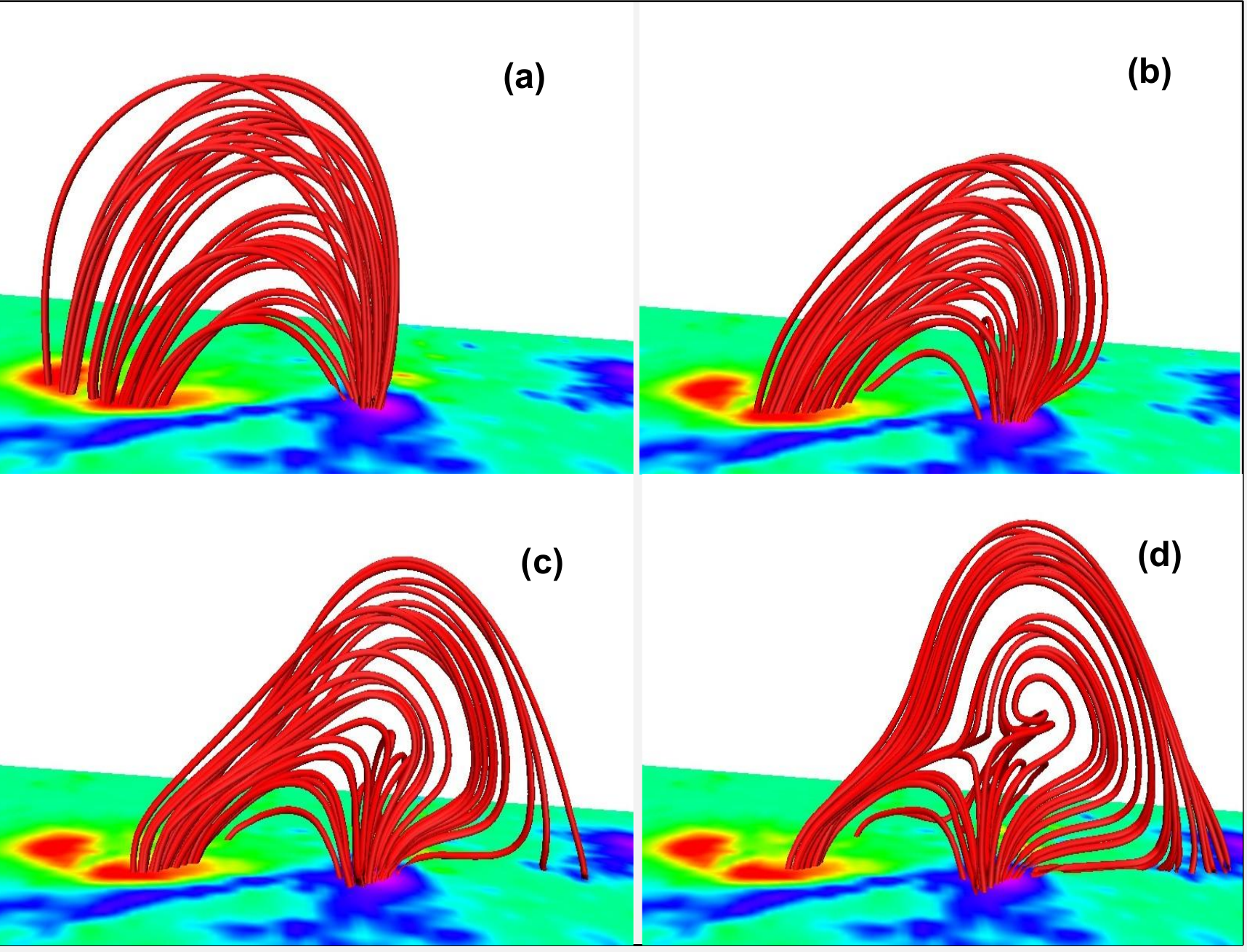}
    \caption{}
    %\label{flow1b}
%\quad    
  \end{subfigure}
  \begin{subfigure}[]{0.42\textwidth}
    \centering
    \includegraphics[width=0.9\linewidth]{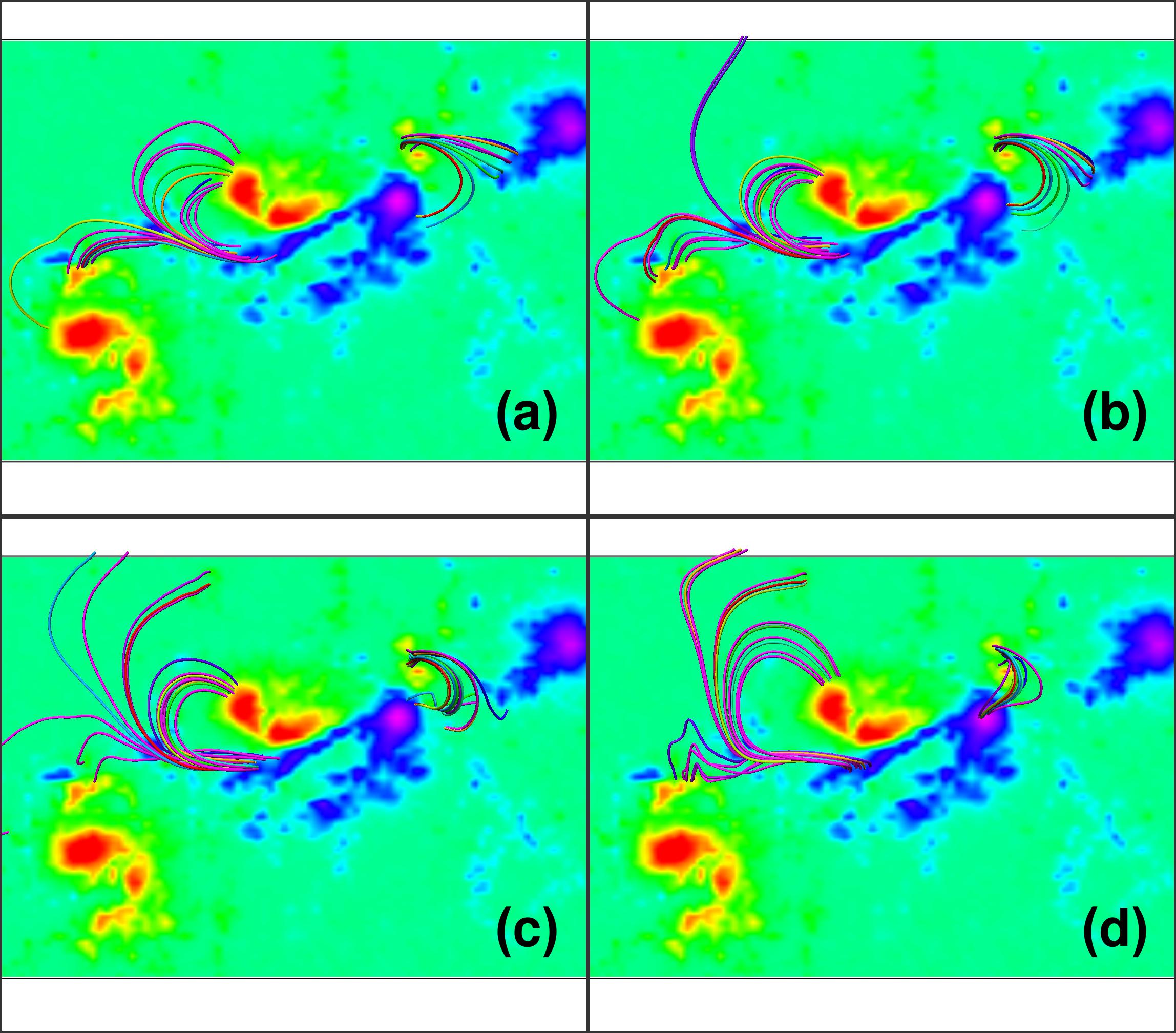}
    \caption{}
    %\label{flow1a}
  \end{subfigure}
    \caption{(a) evolution of MFLs at t=60, 1400, 2100, 2800 depicting the formation of flux rope. (b) flipping of MFLs observed near the qsls at time t=1, 20, 40, 60.}
  \label{f:obs}
\end{figure}

\section{Simulation results}
The plasma is idealized as incompressible, viscous, thermally homogeneous and having perfect electrical conductivity. The MRs, here, are simulated in the sense of Implicit Large Eddy Simulations; induced by a residual dissipation which is generated when scales get under-resolved. The panel (a) Figure (2) plots a MFL system having QSLs marked by bifurcating field lines. Subsequent evolution of the MFLs are documented in the panels (a) to (b). The  MFLs are readily seen to shift their footpoint connectivity from the left of the major bifurcation to its right evident by a temporal increase in number of MFLs on the right. Such slipping reconnections are known to onset blowout jets, which are observed in the collocated region and corroborates to the efficacy of the simulation. However, the post-flare CME indicates the possible presence of a flux-rope which, are not captured at the simulation and is left as a future work.

\end{document}